\documentclass{article}
\usepackage{amsmath,amsfonts,amssymb}
\usepackage{txfonts}
\usepackage[T1]{fontenc}
\usepackage{color}
\newcommand{\red}[1]{\textcolor{black}{#1}}
\newcommand{\blue}[1]{\textcolor{black}{#1}}

\begin{document}
\def\bd{\begin{displaymath}}\def\ed{\end{displaymath}}
\def\be{\begin{equation}}\def\ee{\end{equation}}
\def\bea{\begin{eqnarray}}\def\eea{\end{eqnarray}}
\def\ba{\begin{array}}\def\ea{\end{array}}
\def\nn{\nonumber}\def\lb{\label}\def\bb{\bibitem}

\def\a{\alpha}\def\b{\beta}\def\c{\chi}\def\d{\delta}\def\e{\epsilon}
\def\f{\phi}\def\g{\gamma}\def\h{\theta}\def\i{\iota}\def\j{\vartheta}
\def\k{\kappa}\def\l{\lambda}\def\m{\mu}\def\n{\nu}\def\o{\omega}\def
\p{\pi}\def\q{\psi}\def\r{\rho}\def\s{\sigma}\def\t{\tau}\def\u{\upsilon}
\def\V{\varphi}\def\w{\varpi}\def\y{\eta}\def\x{\xi}\def\z{\zeta}
\def\E{\varepsilon}

\def\D{\Delta}\def\F{\Phi}\def\G{\Gamma}\def\H{\Theta}\def\L{\Lambda}
\def\O{\Omega}\def\P{\Pi}\def\Q{\Psi}\def\S{\Sigma}\def\U{\Upsilon}\def\X{\Xi}

\def\lie{{\cal L}}\def\de{\partial}\def\na{\nabla}\def\per{\times}
\def\inf{\infty}\def\id{\equiv}\def\mo{{-1}}\def\ha{{1\over 2}}
\def\qu{{1\over 4}}\def\pro{\propto}\def\app{\approx}
\def\we{\wedge}\def\di{{\rm d}}\def\Di{{\rm D}}
\def\lra{\leftrightarrow}\def\bdot{\!\cdot\!}

\def\Ei{{\rm Ei}}\def\li{{\rm li}}\def\const{{\rm const}}\def\ex{{\rm e}}
\def\arcsh{{\rm arcsinh}}\def\arcch{{\rm arccosh}}
\def\arcth{{\rm arctanh}}\def\arccth{{\rm arccoth}}
\def\diag{{\rm diag}}

\def\ad{{\rm ad}}

\def\gmn{g_{\m\n}}\def\ep{\e_{\m\n}}\def\ghmn{\hat g_{\m\n}}\def\mn{{\mu\nu}}
\def\dix{\int d^2x\ \sqrt{-g}\ }\def\ds{ds^2=}\def\sg{\sqrt{-g}}
\def\dhx{\int d^2x\ \sqrt{-\hat g}\ }\def\dex{\int d^2x\ e\ }
\def\sn{\mathop{\rm sn}\nolimits}\def\cn{\mathop{\rm cn}\nolimits}
\def\dn{\mathop{\rm dn}\nolimits}

\def\tors#1#2#3{T_{#1#2#3}}\def\curv#1#2#3#4{R_{#1#2#3#4}}
\def\af{asymptotically flat }\def\hd{higher derivative }\def\st{spacetime }
\def\fe{field equations }\def\bh{black hole }\def\as{asymptotically }
\def\eqs{equations }\def\eom{equations of motion }\def\trans{transformations }
\def\tran{transformation }\def\ther{thermodynamical }\def\coo{coordinates }
\def\bg{background }\def\gs{ground state }\def\bhs{black holes }
\def\sc{semiclassical }\def\hr{Hawking radiation }\def\sing{singularity }
\def\ct{conformal transformation }\def\cc{coupling constant }
\def\crel{commutation relations }\def\tl{transformation law }
\def\ns{naked singularity }\def\gi{gravitational instanton }
\def\rep{representation }\def\gt{gauge transformation }
\def\cco{cosmological constant }\def\em{electromagnetic }
\def\ssy{spherically symmetric }\def\cf{conformally flat }
\def\cur{curvature }\def\tor{torsion }\def\ms{maximally symmetric }
\def\coot{coordinate transformation }\def\diff{diffeomorphisms }
\def\gct{general coordinate transformations }\def\gts{gauge transformations }
\def\pb{Poisson brackets }\def\db{Dirac brackets }\def\ham{Hamiltonian }
\def\cd{covariant derivative }\def\dof{degrees of freedom }
\def\hdim{higher dimensional }\def\ldim{lower dimensional }
\def\SR{special relativity }
\def\dys{dynamical system }\def\cps{critical points }\def\dim{dimensional }
\def\sch{Schwarzschild }\def\min{Minkowski }\def\ads{anti-de Sitter }
\def\RN{Reissner-Nordstr\"om }\def\RC{Riemann-Cartan }\def\poi{Poincar\'e }
\def\KK{Kaluza-Klein }\def\pds{pro-de Sitter }\def\des{de Sitter }
\def\BR{Bertotti-Robinson }\def\MP{Majumdar-Papapetrou }
\def\GR{general relativity }\def\GB{Gauss-Bonnet }\def\CS{Chern-Simons }
\def\EH{Einstein-Hilbert }\def\EPG{extended \poi group }
\def\JT{Jackiw-Teitelboim }\def \schr{Schr\"odinger }
\def\dpa{deformed \poi algebra }\def\psm{Poisson sigma model }
\def\td{two-dimensional }\def\trd{three-dimensional }
\def\lt{Lorentz transformations }\def\com{center of mass }
\def\ab{asymptotic behavior}
\def\cor{commutation relations }\def\up{uncertainty principle }
\def\ev{expectation value }\def\bc{boundary conditions }
\def\tran{transformation }\def\ie{i.e.\ }

\def\PL#1{Phys.\ Lett.\ {\bf#1}}\def\CMP#1{Commun.\ Math.\ Phys.\ {\bf#1}}
\def\PRL#1{Phys.\ Rev.\ Lett.\ {\bf#1}}\def\AP#1#2{Ann.\ Phys.\ (#1) {\bf#2}}
\def\PR#1{Phys.\ Rev.\ {\bf#1}}\def\CQG#1{Class.\ Quantum Grav.\ {\bf#1}}
\def\NP#1{Nucl.\ Phys.\ {\bf#1}}\def\GRG#1{Gen.\ Relativ.\ Grav.\ {\bf#1}}
\def\JMP#1{J.\ Math.\ Phys.\ {\bf#1}}\def\PTP#1{Prog.\ Theor.\ Phys.\ {\bf#1}}
\def\PRS#1{Proc.\ R. Soc.\ Lond.\ {\bf#1}}\def\NC#1{Nuovo Cimento {\bf#1}}
\def\JoP#1{J.\ Phys.\ {\bf#1}} \def\IJMP#1{Int.\ J. Mod.\ Phys.\ {\bf #1}}
\def\MPL#1{Mod.\ Phys.\ Lett.\ {\bf #1}} \def\EL#1{Europhys.\ Lett.\ {\bf #1}}
\def\AIHP#1{Ann.\ Inst.\ H. Poincar\'e {\bf#1}}\def\PRep#1{Phys.\ Rep.\ {\bf#1}}
\def\AoP#1{Ann.\ Phys.\ {\bf#1}}\def\AoM#1{Ann.\ Math.\ {\bf#1}}
\def\JHEP#1{JHEP\ {\bf#1}}\def\JCAP#1{JCAP\ {\bf#1}}
\def\RMP#1{Rev.\ Mod.\ Phys.\ {\bf#1}}\def\AdP#1{Annalen Phys.\ {\bf#1}}
\def\grq#1{{\tt gr-qc/#1}}\def\hep#1{{\tt hep-th/#1}}\def\arx#1{{\tt arXiv:#1}}
\def\EPJ#1{Eur.\ Phys.\ J.\ {\bf#1}}

\def\xb{\bar x}\def\f{\varphi}
\def\cH{{\cal H}}\def\cA{{\cal A}}\def\cD{{\cal D}}\def\cP{{\cal P}}\def\cC{{\cal C}}\def\cF{{\cal F}}\def\cI{{\cal I}}
\def\cJ{{\cal J}}\def\cT{{\cal T}}\def\cK{{\cal K}}\def\cG{{\cal G}}\def\cQ{{\cal Q}}\def\cU{{\cal U}}
\def\rhd{\triangleright}\def\ot{\otimes}\def\op{\oplus}\def\Ex{{\rm exp}}

\def\ssb{\left(s_1+\frac{s_2}2\right)}
\begin{titlepage}
\pagenumbering{gobble}
\title{Snyder-type spacetimes, twisted Poincar\'e algebra and addition of momenta}
\vskip80pt
\author{S. Meljanac, D. Meljanac,\\
\small{Rudjer Bo\v skovi\'c Institute, Bijeni\v cka cesta 54, 10002 Zagreb, Croatia,}\\
\\
S. Mignemi and R. \v Strajn\\
\small{Dipartimento di Matematica e Informatica, Universit\`a di Cagliari,}\\
\small{viale Merello 92, 09123 Cagliari, Italy,}\\
\small{and INFN, Sezione di Cagliari, Cittadella Universitaria, 09042 Monserrato, Italy}}
\date{}
\maketitle
\vskip60pt
\begin{abstract}
We discuss a generalisation of the Snyder model \blue{compatible with undeformed Lorentz symmetries, which we describe
in terms of a large class of deformations of the Heisenberg algebra.}
The corresponding deformed addition of momenta, the twist and the $R$-matrix are calculated to first order in
the deformation parameters for all models. In the particular case of the Snyder realisation, \blue{an analytic}
formula for the twist is obtained.
\end{abstract}

\end{titlepage}
\pagenumbering{arabic}

\section{Introduction}
In his seminal paper \cite{Snyder}, Snyder observed that, assuming a noncommutative structure of spacetime,
and hence a deformation of the Heisenberg algebra, it is possible to define a discrete
spacetime without breaking the Lorentz invariance. In this way the short-distance behavior of quantum field
theory can be improved, possibly avoiding ultraviolet divergences.

More recently, noncommutative geometry has become an important field of research \cite{DFR}.
New models have been introduced, as for example the Moyal plane \cite{Moyal} and $\k$-Minkowski
geometry \cite{kappa}, and the formalism of Hopf algebras has been applied to their study \cite{Majid}.
However, contrary to Snyder's, these models either break or deform the action of the Lorentz group on spacetime.

It is therefore interesting to investigate the Snyder model from the point of view of noncommutative geometry.
The Hopf algebra associated with the Snyder model has been studied in a series of papers \cite{BM1,BM2,kappaS},
where the model has been generalised and the star product, coproduct and antipodes have been calculated using the
method of realisations. A different approach was used in \cite {GL}, where the Snyder model was considered in
 a geometrical perspective as a coset in momentum space, and results equivalent to those of refs.\ \cite{BM1,BM2}
 were obtained. More recently, in \cite{MMMS} a further generalisation was introduced and
the construction of QFT on Snyder spacetime was undertaken.

However, some basic properties of the Hopf algebra formalism for Snyder spaces have not yet been investigated:
for example the twist and the related $R$-matrix have not been explicitly calculated, although they have proven
to be very useful tools, especially in the construction of a QFT. In particular, the knowledge of the $R$-matrix
is useful for the definition of a twisted statistics in QFT.
Actually, some difficulties arise because the coproduct in Snyder spaces
is non-coassociative, so that the twist will not satisfy the cocycle condition for the Hopf algebra.

From a different point of view,  phenomenological aspects of the Snyder model have been investigated in classical
and quantum physics, especially in the nonrelativistic 3D limit \cite{CM,mi,LS}. The
most interesting results in this context are the clarification of its lattice-like properties, leading to deformed
uncertainty relations, and the study of the corrections induced on the energy spectrum of some simple physical systems.

In this paper, we extend previous investigations on the noncommutative geometry of the generalised Snyder models, by
calculating the twist and the $R$-matrix to first order in the deformation parameter in the general
case. We also obtain the expression of the twist for the so-called Snyder realisation, introduced in the
original paper \cite{Snyder}.

We note that our results could be rephrased using the formalism of Hopf algebroids
\red{\cite{JMS, Lu, Xu, PLA2013, SIGMA2014, PLB2015, LMP2017, MS1605, JPA2017}}, which is for some
aspects more suitable for the description of the Snyder models than the usual one based on Hopf algebras, since it
deals with the full phase space; however we leave this subject to future investigations.

\section{Snyder space and its generalisation}
Generalised Snyder spaces are a deformation of ordinary phase space, generated by noncommutative \coo $\xb_\m$ and
momenta $p_\m$ that span a deformed Heisenberg algebra $\bar\cH(\bar x,p)$,
\be\lb{snal}
[\xb_\m,\xb_\n]=i\b M_\mn\,\q(\b p^2),\qquad[p_\m,p_\n]=0,\qquad[p_\m,\xb_\n]=-i\f_\mn(\b p^2),
\ee
together with Lorentz generators $M_\mn$ that satisfy the standard relations
\bd
[M_\mn,M_{\r\s}]=i\big(\y_{\m\r}M_{\n\s}-\y_{\m\s}M_{\n\r}+\y_{\n\r}M_{\m\s}-\y_{\n\s}M_{\m\r}\big),
\ed
\be\lb{snal2}
[M_\mn,p_\l]=i\left(\y_{\m\l}p_\n-\y_{\l\n}p_\m\right),\qquad[M_\mn,\xb_\l]=i\left(\y_{\m\l}\xb_\n-\y_{\n\l}\xb_\m\right),
\ee
where the functions $\q(\b p^2)$ and $\f_\mn(\b p^2)$ are constrained so that the Jacobi identities hold,
$\b$ is a constant of the order of $1/M_{Pl}^2$, and $\y_\mn=$\ diag $(-1,1,1,1)$.
The \cor (\ref{snal})-(\ref{snal2}) generalise those originally investigated in \cite{Snyder}, that are recovered for $\q=$ const.

We recall that in its undeformed version, the Heisenberg algebra $\cH(x,p)$ is generated by commutative
\coo $x_\m$ and momenta $p_\m$, satisfying
\be
[x_\m,x_\n]=[p_\m,p_\n]=0,\qquad[p_\m,x_\n]=-i\y_\mn.
\ee
The action of $x_\m$ and $p_\m$ on functions $f(x)$ belonging to the enveloping algebra $\cA$ generated by the $x_\m$
is defined as
\be
x_\m\rhd f(x)=x_\m f(x),\qquad p_\m\rhd f(x)=-i{\de f(x)\over \de x^\m}.
\ee

The noncommutative \coo $\xb_\m$ and the Lorentz generators $M_\mn$ in \eqref{snal}-\eqref{snal2} can be expressed in terms of
commutative \coo $x_\m$ and momenta $p_\m$ as \cite{BM1,BM2}
\be\lb{real}
\xb_\m=x_\m \f_1(\b p^2)+\b \,x\bdot p\,p_\m\f_2(\b p^2)+\b p_\m\c(\b p^2),
\ee
\be\lb{M}
 M_\mn=x_\m p_\n-x_\n p_\m.
\ee
\blue{Note that in the Snyder algebra the generators $M_\mn$ are in principle independent of $x_\m$ and $p_\m$, thus the condition \eqref{M}
characterises a particular representation of the model.}
We also remark that the function $\c$ does not appear in the defining relations \eqref{snal}-\eqref{snal2}, but takes into account the ambiguities
arising from the ordering of the operators $x_\mu$ and $p_\mu$ in equation \eqref{real}.

In terms of the realisation \eqref{real}, the functions $\f_\mn$ in (\ref{snal}) read
\be\lb{phi}
\f_\mn=\y_\mn\f_1+\b p_\m p_\n\f_2,
\ee
while the Jacobi identities are satisfied if
\be\lb{jac}
\q =-2\f_1\f'_1+\f_1\f_2-2\b p^2\f_1'\f_2,
\ee
where the prime denotes a derivative with respect to $\b p^2$. In particular, the function $\psi$
does not depend on the function $\chi$.

From (\ref{jac}) it follows that the \coo $\xb_\m$ are commutative for $\f_2={2\f_1'\f_1\over\f_1-2\b p^2\f_1'}$, and
correspond to Snyder space for $\f_2={1+2\f_1'\f_1\over\f_1-2\b p^2\f_1'}$.
In particular, the Snyder realisation \cite{Snyder} is recovered for $\f_1=\f_2=1$, and the Maggiore \cite{Mag} realisation for
$\f_1=\sqrt{1-\b p^2}$, $\f_2=0$ \cite{BM1,BM2}. There is also another interesting exact realisation of Snyder space
for $\q=s=$ const,
given by
\be \label{conf}
\xb_\m=x_\m+{\b s\over4}\,K_\m,
\ee
\red{which corresponds to $\c=0$, and}
where $K_\m=x_\m p^2-2x\bdot p\,p_\m\ $ are the generators of conformal transformations in momentum space,
with $[K_\m,K_\n]=0$.
The algebra (\ref{snal}) unifies commutative space, $\q=0$, and Snyder space, $\q=1$. Since the Lorentz transformations
 are not deformed, the Casimir operator of the algebra (\ref{snal})-(\ref{snal2}) is $\cC=p^2$.

 The Hopf algebra associated with these spaces can be investigated using the formalism introduced in
 refs.\ \red{\cite{BM2, kappaS, Svrtan, KM, KMSS}} and \red{generalised} in \cite{MM}, to which we refer for more details.
It turns out that the generalised addition of momenta $k_\m$ and $q_\m$ is given by \cite{BM2,kappaS,JMP}
\be
k_\m\oplus q_\m=\cD_\m(k,q),\qquad{\rm with}\quad\cD_\m(k,0)=k_\m,\quad\cD_\m(0,q)=q_\m,
\ee
where $k,q \in M_{1,3}$.
The function $\cD_\m(k,q)$ can be calculated in terms of $\f_\mn$ as
\be\lb{DPK}
\cD_\m(k,q)=\cP_\m(\cK^\mo(k),q),
\ee
\red{where we have introduced the function $\cK_\m(k)=\cP_\m(k,0)$ and its inverse  $\cK^{-1}_\mu(k)$,
such that $\cK^{-1}_\m(\cK(k))=k_\m$. The function $\cP_\m(\l k,q)$} satisfies the differential equation
\be\lb{diffe}
{d\cP_\m(\l k,q)\over d\l}=k_\a\f_\m^{\ \a}\Big(\cP(\l k,q)\Big), \qquad \red{\l\in\mathbb R,}
\ee
with
\be
\cP_\m(k,0)=\cK_\m(k),\qquad\cP_\m(0,q)=q_\m,
\ee
From \eqref{diffe} and \eqref{phi} it follows that $\cP_\m(k,q)$ and hence $\cD_\m(k,q)$ do not depend
on the function $\c$ in \eqref{real}.

It can be shown that \cite{Svrtan, KM, KMSS}
\be
e^{ik\cdot\xb}\rhd e^{iq\cdot x}=e^{i\cP(k,q)\cdot x+i\cQ(k,q)},
\ee
where $\cQ(k,q)$ satisfies the differential equation
\be
{d\cQ(\l k,q)\over d\l}=k_\a\c^\a\Big(\cP(\l k,q)\Big),
\ee
with  $\cQ(0,q)=0$ and $\c^\a\id p^\a\c(\b p^2)$.

Calculating the star product of two plane waves one then obtains
\be \lb{starexp}
e^{ik\cdot x}\star e^{iq\cdot x} = e^{i\cD(k,q)\cdot x+i\cG(k,q)},
\ee
with
\be
\cG(k,q)=\cQ(\cK^\mo(k),q)-\cQ(\cK^\mo(k),0).
\ee
Note that $\cG$ vanishes if $\c(k)=0$.

The algebra $\cA$\red{, generated by the coordinates $x_\m$,} can be extended to the algebra $\cU$ generated by the $x_\m$ and the $p_\m$,
symbolically indicated as $\cU=\cA\,\cT$, where $\cT$ is the algebra generated by the $p_\m$ \red{\cite{JMS, JMP}}.
The coproduct for the momenta $\D p_\m$, is obtained from $\cD_\m(k,q)$ as
\be\lb{cop}
\D p_\m=\cD_\m(p\ot1,1\ot p).
\ee
Notice that the previous definitions imply that the addition of momenta and the coproduct do not depend on $\c(\b p^2)$.

\red{From \eqref{starexp} and the coproduct \eqref{cop}} one can then define the twist $\cF$, such that $\D h=\cF\D_0h\cF^\mo$ for any $h\in\cU$,
as  \cite{JMS,GGH,MSS}
\be\lb{twist}
\cF^\mo=:\Ex\big[i(1\ot x_\a)(\D-\D_0)p_\a+\red{i}\,\mathcal{G}(p\otimes 1, 1\otimes p)\big]:\ ,
\ee
where $\D_0p_\m=p_\m\ot1+1\ot p_\m$, and $:\ \ :$ denotes normal ordering in which the \coo $x_\a$ stand on
the left of the momenta $p_\a$.

The star product $f\star g$ can be defined as
\be \lb{fstarg}
\big(f\star g\big)(x)=m\Big(\cF^\mo(\rhd\ot\rhd)(f\ot g)\Big),\qquad f,g\in\cA,
\ee
with $m: \cA \otimes \cH \rightarrow \cH$ the multiplication map of $\cA$.

The relation (\ref{real}) between $\xb_\m$ and $x_\m$ can also be written in terms of the twist as
\be
\xb_\m=m\Big(\cF^\mo(\rhd\ot1)(x_\m\ot1)\Big)=x_\a\f_{\ \m}^\a(p)+\b p_\m\c(p).
\ee
It follows for consistency that
\be\lb{ptwist}
\D p_\m=\cF(\D_0p_\m)\cF^\mo, \qquad \Delta_0 p_\mu =p_\mu \otimes 1 + 1 \otimes p_\mu
\ee
in accordance with (\ref{cop}).

The coproducts of momenta are found for special cases in \cite{BM2}: for the Snyder realisation
\begin{equation} \label{coprod}
\Delta p_{\mu} = \frac{1}{1-\b p_{\alpha}\otimes p^{\alpha}} \left(p_{\mu} \otimes 1 - \frac{\b}
{1+\sqrt{1+\b p^2}}\, p_{\mu}p_{\alpha} \otimes p^{\alpha} + \sqrt{1+\b p^2} \otimes p_{\mu} \right),
\end{equation}
while for the Maggiore realisation
\begin{equation}\label{magg}
\Delta p_\mu =p_\mu \otimes \sqrt{1-\b p^2} -\frac{\b}{1+\sqrt{1-\b p^2}}\, p_\mu p_\alpha \otimes p^\alpha
+1\otimes p_\mu.
\end{equation}
The coproducts of the Lorentz generators are instead
\be \label{copMprekoF}
\D M_\mn=\cF(\D_0M_\mn)\cF^\mo,\qquad\D_0M_\mn=M_\mn\ot1+1\ot M_\mn.
\ee
\red{Generally,} because of the commutation relations \eqref{snal2}, the coproduct of $M_{\mu\nu}$ will be trivial, i.e. $\Delta M_{\mu\nu} =\Delta_0 M_{\mu\nu}$ \cite{BM2}.

We recall that also the antipodes for Snyder space are trivial \cite{BM2},
\be
S(p_\m)=-p_\m,\qquad S(M_\mn)=-M_\mn,
\ee

\section{First order expansion}
The study of the general form of the deformed Heisenberg algebra \eqref{snal} is difficult, however one can study it perturbatively,
by expanding the realisation \eqref{real} of the noncommutative \coo in powers of $\b$, namely,
\be \label{1redx}
\xb_\m=x_\m+\b\,(s_1x_\m p^2+s_2x\bdot p\,p_\m+cp_\m)+O(\b^2),
\ee
with parameters $s_1$, $s_2$, $c$\red{, such that $\f_1(\b p^2)=1+s_1\b p^2 + O(\b^2)$, $\f_2(\b p^2) = s_2 + O(\b)$ and $\c(\b p^2)=c+O(\b)$}.
Hence, the \cor do not depend on the parameter $c$ and to first order are given by
\be\lb{cor}
[\xb_\m,\xb_\n]=i\b sM_\mn +O(\beta^2),\qquad[p_\m,\xb_\n]=-i\,\big[\y_\mn(1+\b s_1p^2)+\b s_2p_\m p_\n\big]+O(\beta^2),
\ee
where $s=s_2-2s_1$.

The models of ref.\  \cite{BM1,BM2} are recovered for $s_2=1+2s_1$. Moreover,
for $s_1=0$, $s_2=1$, eqs.\ \eqref{1redx}-\eqref{cor} reproduce the exact Snyder realisation, while for $s_1=-\ha$, $s_2=0$
they give the first-order expansion of the Maggiore realisation. For $s_2=2s_1$, spacetime is commutative to first order
in $\beta$, \blue{although the commutation relations are not canonical,} while for $s_1=-s/4$, $s_2=s/2$, $c=0$ one gets the exact realisation \eqref{conf}.

The first order expression for the function $\mathcal{P}_\mu (k,q)$ is given by
\begin{eqnarray}
&& \mathcal{P}_\mu (k,q)= q_\mu +\int_0^1 d\lambda \left\{ k_\mu +\beta \left[ s_1 k_\mu (\lambda k +q)^2 +
s_2 (\lambda k^2 +k\cdot q) (\lambda k_\mu +q_\mu) \right] \right\} +O(\beta^2) \nonumber \\
&& = k_\mu +q_\mu + \beta \left[ \left(s_1 q^2 + \left( s_1+\frac{s_2}{2} \right) k\cdot q +
\frac{s_1+s_2}{3} k^2 \right) k_\mu + s_2 \left( k\cdot q +\frac{k^2}{2} \right) q_\mu \right]\nonumber\\
&&+O(\beta^2),
\end{eqnarray}
from where it follows that
\begin{equation}\lb{cK1}
\cK^{-1}_\mu (k) =k_\mu -\frac{\beta}{3}(s_1+s_2) k^2 k_\mu +O(\beta^2).
\end{equation}
\red{Note that for $s_1+s_2=0$, $\cK_\mu(k)=\cK^{-1}_\mu(k)=k_\mu$, to the first order in $\b$.}
These results allow us to write down the generalised addition law of the momenta $k_\m$ and $q_\m$ to first order
\be\lb{linadd}
(k\op q)_\m= \mathcal{D}_\mu (k,q)= k_\m+q_\m+\b\left[s_2k\bdot q\,q_\m+s_1q^2k_\m
+\left(s_1+{s_2\over2}\right)k\bdot q\,k_\m+{s_2\over2}k^2q_\m\right]+O(\b^2).
\ee
In particular, for the "conformal" case \eqref{conf} with parameters $s_1=-s/4$, $s_2=s/2$,
\be
(k\op q)_\m=k_\m+q_\m+{\b s\over4}\,\big[2\,k\bdot q\,q_\m-q^2k_\m+k^2q_\m\big]+O(\b^2).
\ee
It is also interesting to remark that for $s_2=2s_1\neq0$, $s=0$, although spacetime is commutative up to the first order in $\beta$,
the addition of momenta is still deformed, \blue{but it is now commutative}
\be
(k\op q)_\m=(q\op k)_\m\ne k_\m+q_\m.
\ee

The Lorentz transformations of momenta are not deformed, and denoting them by $\L(\x,p)$, with $\x$ the rapidity
parameter, the law of addition of momenta implies that
\be
\L(\x,k\op q)=\L(\x_1,k)\op\L(\x_2,q)
\ee
is satisfied for $\x_1=\x_2=\x$. Hence there are no backreaction factors in the sense of ref. \cite{GM,maj}.
This means that in composite systems the boosted momenta of the single particles are independent of the momenta of the
other particles in the system.

The coproduct to \red{the} first order can be read from (\ref{linadd}) and is given by
\begin{equation} \label{coprod1ord}
\Delta p_\mu = \Delta_0 p_\mu + \beta \left[ s_1 p_\mu \otimes p^2 + s_2 p_\alpha \otimes p^\alpha p_\mu +\left( s_1 +
\frac{s_2}{2} \right) p_\mu p_\alpha \otimes p^\alpha + \frac{s_2}{2} p^2 \otimes p_\mu \right] + O(\beta^2).
\end{equation}

The corresponding twist operator $\mathcal{F}^{-1}$ is
\begin{equation} \label{twist1ord}
\mathcal{F}^{-1} = 1\otimes 1 + i(1 \otimes x_\alpha )(\Delta - \Delta_0) p^\alpha + ic\beta p_\alpha \otimes p^\alpha +O(\beta^2).
\end{equation}
\red{or equivalently, in terms of dilatation $D=x\cdot p$ and momenta $p_\a$
\be\begin{split}\lb{twist1ordD}
\cF^{-1}&=1\otimes1 + i\b\left[
s_1 D \otimes p^2 + \frac{s_2}2 p^2 \otimes D + s_2 p_\a \otimes D p^\a
+ \left(s_1+\frac{s_2}2\right)Dp_\a \otimes p^\a
\right]\\
&+ic\b p_\a \otimes p^\a + O(\b^2).
\end{split}
\ee
}
From this one can calculate the coproduct $\Delta M_{\mu \nu}$,  and the antipodes $S(p_\mu)$ and $S(M_{\mu \nu})$ \red{to the first order in $\b$}.
Using the twist \eqref{twist1ord}, \eqref{twist1ordD} to calculate the coproduct of $p_\mu$ as in \eqref{ptwist}, one gets again \eqref{coprod1ord},
the same result as when using the function $\mathcal{D}$, while using \eqref{copMprekoF} to calculate the coproduct of $M_{\mu \nu}$
gives $\Delta M_{\mu \nu} = \Delta_0 M_{\mu \nu} +O(\beta^2)$, \red{which is consistent with the general result $\D M_{\m\n} = \D_0 M_{\m\n}$}.

In general\red{, the} twist \red{\eqref{twist}} will not satisfy the cocycle condition, the star product \red{\eqref{starexp}, \eqref{fstarg}} will
be non-associative \blue{and the coproduct $\Delta p_\mu$ in \eqref{cop} will be non-coassociative.} Let us verify these claims.

\red{The cocycle condition is $(\cF\otimes1)(\D_0\otimes1)\cF=(1\otimes\cF)(1\otimes\D_0)\cF$.
The left hand side calculated for the twist $\cF$ \eqref{twist1ord} to the first order in $\b$ is
\be
\begin{split}
&1\ot1\ot1 + i\b\left(\vphantom\ssb
s_1 p_\a\ot1\ot x^\a p^2 + s_11\ot p_\a\ot x^\a p^2 +
s_2p_\a\ot1\ot x\cdot p p^\a \right. \\&+ s_21\ot p_\a \ot x\cdot p p^\a + \ssb p_\a p_\b \ot 1\ot x^\a p^\b +
\ssb1\ot p_\a p_\b \ot x^\a p^\b \\&+\ssb p_\a \ot p_\b \ot x^\a p^\b +
\ssb p_\a \ot p_\b \ot x^\b p^\a + \frac{s_2}2p^2\ot1\ot x\cdot p\\& +
\frac{s_2}21\ot p^2 \ot x\cdot p + s_2 p_\a \ot p^\a \ot x\cdot p + s_1 p_\a \ot p^2 \ot x^\a +
s_2 p_\a \ot p^\a p_\b \ot x^\b \\&\left.+ \ssb p_\a p_\b \ot p^\a \ot x^\b +
\frac{s_2}2 p^2 \ot p_\a \ot x^\a \vphantom\ssb\right) + O(\b^2),
\end{split}
\ee
while the right hand side is
\be
\begin{split}
&1\ot1\ot1 + i\b\left(\vphantom\ssb s_1 p_\a \ot p^2 \ot x^\a + s_1 p_\a \ot 1 \ot x^\a p^2 + 2s_1 p_\a \ot p_\b \ot x^\a p^\b \right.\\
&+s_2 p_\a \ot p^\a p_\b \ot x^\b + s_2 p_\a \ot p_\b \ot x^\b p^\a + s_2 p_\a \ot 1 \ot x \cdot p  p^\a \\&+
s_2 p_\a \ot p^\a \ot x\cdot p + \ssb p_\a p_\b \ot p^\b \ot x^\a + \ssb p_\a p_\b \ot1\ot x^\a p^\b \\&+
\frac{s_2}2 p^2 \ot p_\a \ot x^\a + \frac{s_2}2 p^2 \ot 1 \ot x\cdot p + s_11\ot p_\a \ot x^\a p^2 + s_21 \ot p_\a \ot x\cdot p p^\a \\
&\left.+ \ssb1\ot p_\a p_\b \ot x^\a p^\b + \frac{s_2}21\ot p^2 \ot x\cdot p \vphantom\ssb\right) + O(\b^2).
\end{split}
\ee}
\red{Generally, the cocycle condition is not satisfied even to the first order in $\b$. It is satisfied only in the special case $s_2=2s_1$,
which corresponds to commutative space.}
\red{The coassociativity condition for the coproduct $\D$ \eqref{cop}, \eqref{coprod1ord} is $(\D\ot1)\D=(1\ot\D)\D$.
In general, also the coassociativity condition is not satisfied even to the first order in $\b$, except in the special case $s_2=2s_1$.}

\red{The associativity condition for the star product \eqref{starexp}, \eqref{fstarg}, \eqref{linadd} is
\begin{equation}
e^{ik_1\cdot x}\star(e^{ik_2\cdot x}\star e^{ik_3\cdot x})=(e^{ik_1\cdot x}\star e^{ik_2\cdot x})\star e^{ik_3\cdot x}.
\end{equation}
The left hand side, calculated to the first order in $\b$ is
\be
\begin{split}
&k_{1\m}+k_{2\m}+k_{3\m}+\b\left( k_{1\m}\left[s_1(k_2^2+k_3^2+2k_2\cdot k_3) +\ssb(k_1\cdot k_2+k_1\cdot k_3)\right]+\right. \\
&k_{2\m}\left[s_1 k_3^2 + \ssb k_2\cdot k_3 + s_2(k_1\cdot k_2 + k_1\cdot k_3) + \frac{s_2}2k_1^2\right]+ \\
&\left.\vphantom\ssb
k_{3\m}\left[s_2 k_2\cdot k_3 + \frac{s_2}2 k_2^2 + s_2(k_1\cdot k_2 + k_1\cdot k_3) + \frac{s_2}2 k_1^2\right]
\right) + O(\b^2),
\end{split}
\ee
while the right hand side reads
\be
\begin{split}
&k_{1\m}+k_{2\m}+k_{3\m}+\b\left(k_{1\m}\left[s_1k_2^2 + \ssb k_1\cdot k_2 + s_1 k_3^2 + \ssb(k_1\cdot k_3 + k_2\cdot k_3)
\right]+\right. \\
&k_{2\m}\left[s_2k_1\cdot k_2 + \frac{s_2}2k_1^2+s_1k_3^2 + \ssb(k_1\cdot k_3 + k_2\cdot k_3) \right]+ \\
&\left. k_{3\m}\left[s_2(k_1\cdot k_3 + k_2\cdot k_3) + \frac{s_2}2(k_1^2+k_2^2+2k_1\cdot k_2) \right]\right) + O(\b^2).
\end{split}
\ee}
\red{It follows that the associativity condition for the star product is again not satisfied even to the first order in $\b$, except in the special case $s_2=2s_1$.
In this case, the star product is commutative and associative, but non-local.}

\blue{Finally, in the commutative case $s_2=2s_1$, it is easily seen from \eqref{coprod1ord} that}
\be\label{tau}
\tilde{\Delta} p_\mu \id \tau_0 \Delta p_\mu \tau_0 =\Delta p_\mu,
\ee
i.e.\ the coproduct  is left-right symmetric, with the
flip operator $\tau_0$ defined in the usual way as
\be
\tau_0 (A\otimes B) = B \otimes A.
\ee
The coproduct is cocommutative and \red{coassociative.}

\red{In the general case $s_2\ne2s_1$, the algebraic structure of the generalised Hopf algebra defined by the
enveloping algebra of Poincar\'e-Weyl algebra, the coproduct $\D$, the antipode $S$ and the counit $\e$ is under investigation.}

The flip operator, $\tau =\mathcal{F} \tau_0 \mathcal{F}^{-1}$, \red{$\t^2=\t_0^2=1\otimes1$} is relevant in the discussion of the twisted statistics of
particles in quantum field theory on noncommutative
spaces \cite{GGH, MSS}. \red{Note that
\be
m\cF^{-1}=m\t_0\cF^{-1}=m\cF^{-1}\t=m\tilde\cF^{-1}\t_0
\ee
where $\tilde\cF=\t_0\cF\t_0$.} Another important operator in this context is the $R$-matrix, which satisfies the relation
$R\,\D p_\m R^\mo=\tilde\D p_\m$ \red{and $\t=R^{-1}\t_0=\t_0R$}.
\red{The $R$-matrix} can be written as
\be
R= \tilde{\mathcal{F}} \mathcal{F}^{-1} = 1\otimes 1 + R_{cl} + O(\beta^2),
\ee
\red{and it should lead to a generalization of the triangularity condition.} 
The classical $R$-matrix $R_{cl}$  is
\be \lb{rcl}
R_{cl}= (x_\alpha \otimes 1) (\tilde{\Delta} - \Delta_0) p^\alpha - (1\otimes x_\alpha) (\Delta -\Delta_0) p^\alpha,
\ee
where $\Delta p_\mu$ is given in \eqref{coprod1ord}. \red{From \eqref{twist1ordD} and \eqref{rcl}, we can write
\be \lb{rclD}
R_{cl}=\b\left(s_1-\frac{s_2}2\right)
[D\otimes p^2 - p^2\otimes D + (D\otimes1 - 1\otimes D)p_\a\otimes p^\a],
\ee
or equivalently $R_{cl}=\b\left(s_1-\frac{s_2}2\right)(M_{\a\b}p_\b\otimes p_\a - p_\a \otimes M_{\a\b}p_\b)$. For $s_1=\frac{s_2}2$ it follows that $R_{cl}=0$.}
For commutative spaces, for which \eqref{tau} holds, $R_{cl}$ is given by
\be
R_{cl} = (x_\alpha \otimes 1 - 1\otimes x_\alpha) (\Delta - \Delta_0) p^\alpha \in \mathcal{I}_0.
\ee
where $\cI_0$ is the right ideal of $\cU$ with the property $m\Big(\cI_0\rhd(f\ot g)\Big)=0$.

\red{
\subsection{Hopf algebroid approach}
The Hopf algebroid structure was introduced in \cite{Lu, Xu}. We point out that the twist $\cF^{-1}$ in \eqref{twist}, \eqref{twist1ord},
as well as in the classical $R$-matrix $R_{cl}$ in \eqref{rcl}, \eqref{rclD} are obtained in the Hopf algebroid approach \cite{JMS, PLA2013, SIGMA2014},
where the set of generators $x_\m$ and $p_\m$ defines the basis of the Heisenberg Hopf algebroid \cite{PLB2015, LMP2017}.
In the general case $s_2\ne2s_1$, the algebraic structure of the generalised Hopf algebroid is currently under investigation.
}

\red{
An important result is that the twist $\cF^{-1}$ \eqref{twist} in the case $\cG=0$ (i.e. $\c(p)=0$) is identical \cite{MS1605, JPA2017} to
\be
\cF^{-1}=
e^{-ip_\a\otimes x^\a}e^{i\cK^{-1}_\g(p)\otimes\xb^\g}
=e^{-ip_\a\otimes x^\a}e^{i\cK^{-1}_\g(p)\otimes x^\b \f_\b{}^\g(p)},
\ee
where $\cK^{-1}_\g(p)$ is defined after \eqref{DPK}.
}

\red{
To the first order in $\b$, $\cK^{-1}_\g(p)$ is given in \eqref{cK1}. For example, in the Snyder case, $s_1=0$, $s_2=1$ and
\be
\cK^{-1}_\g(p)=p_\g-\frac\b3p^2p_\g+O(\b^2).
\ee
In the Maggiore realisation, $s_1=-\frac12$, $s_2=0$ and
\be
\cK^{-1}_\g(p)=p_\g+\frac\b6p^2p_\g+O(\b^2).
\ee
}

\section{Twist for the Snyder realisation}
In this section,
we construct the exact twist operator for the Snyder space using the perturbative approach introduced in \cite{JMS},
by expanding \eqref{ptwist} in powers of $\b$.
We first consider the special case of the Snyder realisation $\f_1=\f_2=1$, $\c=0$, for which
\begin{equation}
\xb_{\mu}=x_{\mu}+\b\,x\bdot p\,p_{\mu}.
\end{equation}

The coproduct of the momenta is given by (\ref{coprod}).
We expand it with respect to the deformation parameter $\b$ as $\Delta p_\mu = \sum^{\infty}_{k=0} \Delta_k p_\mu$, with $\Delta_k p_\mu \propto \b^k$
\begin{eqnarray}
\Delta p_\mu &=& p_\mu \otimes 1+1 \otimes p_\mu +\b\, \Biggl( \frac{1}{2} p_\mu p_\alpha \otimes p^\alpha
+ p_\alpha \otimes p^\alpha p_\mu + \frac{1}{2} p^2 \otimes p_\mu \Biggr) \nonumber \\
&+& \b^2\, \Biggl( \frac{1}{2} p_\mu p_\alpha p_\beta \otimes p^\alpha p^\beta + p_\alpha p_\beta \otimes p^\alpha p^\beta p_\mu
+\frac{1}{8} p_\mu p_\alpha p^2 \otimes p^\alpha -\frac{1}{8} p^4 \otimes p_\mu \nonumber \\
&+& \frac{1}{2} p_\alpha p^2 \otimes p^\alpha p_\mu \Biggr) + \b^3\, \Biggl( \frac{1}{2} p_\mu p_\alpha p_\beta p_\gamma \otimes p^\alpha p^\beta p^\gamma
+ p_\alpha p_\beta p_\gamma \otimes p^\alpha p^\beta p^\gamma p_\mu \nonumber \\
&-& \frac{1}{16} p_\mu p_\alpha p^4 \otimes p^\alpha +\frac{1}{8} p_\mu p_\alpha p_\beta p^2 \otimes p^\alpha p^\beta +
\frac{1}{16} p^6 \otimes p_\mu -\frac{1}{8} p_\alpha p^4 \otimes p^\alpha p_\mu \nonumber \\
&+& \frac{1}{2} p_\alpha p_\beta p^2 \otimes p^\alpha p^\beta p_\mu \Biggr) +O(\b^4)
\end{eqnarray}
and we look for the twist operator in the form
\begin{equation}
\mathcal{F}= e^{f_1 +f_2 +f_3 +...},
\end{equation}
where $f_k\propto \b^k$. From \eqref{ptwist} we obtain the equations satisfied by the $f_k$ order by order,
\begin{eqnarray}
 \left[f_{1},\Delta_{0} p_{\mu}\right] &=& \Delta_{1} p_{\mu}, \label{anf1} \\
 \left[ f_{2},\Delta_{0} p_{\mu}\right] &=& \Delta_{2} p_{\mu}- \frac{1}{2} \left[ f_{1}, \left[ f_{1}, \Delta_{0} p_{\mu} \right] \right], \label{anf2} \\
 \left[ f_3 ,\Delta_0 p_\mu \right] &=& \Delta_3 p_\mu -\frac{1}{2}\left( \left[ f_1 ,\left[ f_2 ,\Delta_0 p_\mu \right] \right]+
 \left[ f_2 ,\left[ f_1 ,\Delta_0 p_\mu \right] \right] \right)\nonumber \\
&&-\frac{1}{3!} \left[ f_1 ,\left[ f_1 ,\left[ f_1 ,\Delta_0 p_\mu \right] \right] \right],
\end{eqnarray}
and so on.
To calculate $f_1$ we write down the ansatz
\begin{equation} \nonumber
f_1= \b\,\big( \alpha_1 p^2 \otimes x\bdot p + \alpha_2 p_\alpha p_\beta \otimes x^\alpha p^\beta + \alpha_3 p_\alpha \otimes x\bdot p\, p^\alpha +
\alpha_4 p_\alpha \otimes x^{\alpha} p^2\big)
\end{equation}
and insert it into \eqref{anf1} to determine the unknown coefficients $\alpha_i$. The resulting expression for $f_1$ is
\begin{equation} \label{f1}
f_1= -i\b\,\left( \frac{1}{2} p^2\otimes x\bdot p+ \frac{1}{2} p_\alpha p_\beta \otimes x^\alpha p^\beta + p_\alpha \otimes x\bdot p\,p^\alpha \right).
\end{equation}
Inserting this and the ansatz
\begin{eqnarray}
f_2 &=& \b^2 \big(\alpha_1 p^4 \otimes x\bdot p + \alpha_2 p_\alpha p_\beta p^2 \otimes x^\alpha p^\beta + \alpha_3 p_\alpha p^2 \otimes x\bdot p\, p^\alpha \nonumber \\
&+& \alpha_4 p_\alpha p^2 \otimes x^\alpha p^2 + \alpha_5 p_\alpha p_\beta p_\gamma \otimes x^\alpha p^\beta p^\gamma \big), \nonumber
\end{eqnarray}
into \eqref{anf2}, we find
\begin{equation} \label{f2}
f_2= i\,\frac{\b^2}{2} \left( \frac{1}{2} p^4 \otimes x\bdot p +\frac{1}{2} p_\alpha p_\beta p^2 \otimes x^\alpha p^\beta + p_\alpha p^2 \otimes x\bdot p\, p^\alpha \right).
\end{equation}
An analogous procedure to third order gives
\begin{equation}
f_3= -i\,\frac{\b^3}{3}\left( \frac{1}{2} p^6 \otimes x\bdot p+ \frac{1}{2} p_\alpha p_\beta p^4 \otimes x^\alpha p^\beta
+ p_\alpha p^4 \otimes x\bdot p\, p^\alpha \right).
\end{equation}

\red{From the results for $f_1$, $f_2$, $f_3$,... we conjecture that the twist $\cF$ can be written as}
\begin{equation}\lb{fsn}
\mathcal{F}= \exp \left\{ -i\left( \frac{1}{2} p^2 \otimes x\bdot p +\frac{1}{2} p_{\alpha} p_{\beta} \otimes x^{\alpha} p^{\beta}
+ p_{\alpha} \otimes x\bdot p\, p^{\alpha} \right) \left( \frac{\red{\ln}(1+\b p^2)}{p^2} \otimes 1 \right) \right\}.
\end{equation}
One can check that \eqref{fsn} gives the correct twist for the Snyder space by calculating
\begin{equation}
m\Big(\cF^\mo(\rhd\ot1)(x_\m\ot1)\Big) =x_\m+\b\,x\bdot p\,p_{\mu}.
\end{equation}

An independent verification is to start from (\ref{twist}). We get
\begin{eqnarray}
&& \mathcal{F}^{-1} =\ : \exp\ \Biggl[\frac{i}{1-\b\, p_\alpha \otimes p^\alpha} \Biggl( \frac{\b\sqrt{1+\b p^2}}
{1+\sqrt{1+\b p^2}}\, p^\mu p^\nu \otimes x_\mu p_\nu + \left( \sqrt{1+\b p^2} -1\right) \otimes x\bdot p \nonumber \\
&& \qquad \quad +\b\, p_\alpha \otimes x\bdot p\, p^\alpha \Biggr) \Biggr]:\ , \label{2met}
\end{eqnarray}
which expanded up to second order gives
\begin{eqnarray}
\mathcal{F}^{-1} &=& 1\otimes 1 + i\b\, \Biggl( \frac{1}{2} p^\alpha p^\beta \otimes x_\alpha p_\beta
+\frac{1}{2} p^2 \otimes x\bdot p +p_\alpha \otimes x\bdot p p^\alpha \Biggr) \nonumber \\
&-& \frac{i\b^2}{2} \Biggl( \frac{1}{4} p^4 \otimes x\bdot p - \frac{1}{4} p_\alpha p_\beta p^2 \otimes x^\alpha p^\beta -
p_\alpha p^2 \otimes x\bdot p p^\alpha - p_\alpha p_\beta p_\gamma \otimes x^\alpha p^\beta p^\gamma \nonumber \\
&-& 2p_\alpha p_\beta \otimes x\bdot p p^\alpha p^\beta \Biggr) -\frac{\b^2}{2} \Biggl( \frac{1}{4} p^4
\otimes x^\alpha x\bdot p p_\alpha + \frac{1}{2} p_\alpha p_\beta p^2 \otimes x^\alpha x\bdot p p^\beta \nonumber \\
&+& p_\alpha p^2 \otimes x_\beta x\bdot p p^\beta p^\alpha + \frac{1}{4} p_\alpha p_\beta p_\gamma p_\delta \otimes
x^\alpha x^\beta p^\gamma p^\delta + p_\alpha p_\beta p_\gamma \otimes x^\alpha x\bdot p p^\beta p^\gamma \nonumber \\
&+& p_\alpha p_\beta \otimes x_\gamma x\bdot p p^\gamma p^\alpha p^\beta \Biggr) +O(\beta^3). \label{2metpert}
\end{eqnarray}
The expression in eq.\ \eqref{2metpert} agrees exactly with what one would get from \eqref{f1} and \eqref{f2} using the fact that
$\,\mathcal{F}^{-1} = 1\otimes 1 -f_1 -f_2 +\frac{1}{2} f_1^2+ O(\b^3)$.

\red{As a further check, let us calculate the coproduct $\D p_\m = \cF \D_0 p_\m \cF^{-1}$ with twist $\cF$ given in \eqref{fsn}
\be
\cF\D_0p_\m\cF^{-1} = p_\m\ot1 + \sum_{n=0}^\infty \sum_{k=0}^n
\b^{n-k}\frac{(-1)^{k+n}}{k!}A_{n,k} (p^{2(n-k)}\ot1)\, \ad_{f_1}^k(1\ot p_\m),
\ee
where
\be
A_{n,k}=\sum_{r_1+...+r_k=n} \frac1{r_1r_2...r_k}
\ee
and
\be\begin{split}
\ad_{f_1}^k(1\otimes p_\m)&=\b^k\left[
\sum_{l=0}^k c_{k-l,l}(p_\m (p^2)^{k-l}\ot1)(p_\a \ot p^\a)^l+
\right. \\&\phantom{=\b^k\left[\vphantom{\sum_a^b}\right.}\left.
\sum_{l=0}^k d_{k-l,l}((p^2)^{k-l}\ot p_\m)(p_\a \ot p^\a)^l\right].
\end{split}\ee
The coefficients $c_{k-l,l}$ and $d_{k-l,l}$ satisfy the following recursive relations
\begin{align}
c_{k-l+1,l}&=lc_{k-l,l} + (l-1)c_{k-l+1,l-1} + \frac12 d_{k-l+1,l-1} \\
d_{k-l+1,l}&=\left(l+\frac12\right)d_{k-l,l} + ld_{k-l+1,l-1}
\end{align}
with $c_{0,0}=0$ and $d_{0,0}=1$. Particularly, the coefficients $c_{k,0}$, $c_{k-1,1}$ and $d_{k,0}$ are
\be\lb{ccd}
c_{k,0}=0, \qquad c_{k-1,1}=1-\frac1{2^k}, \qquad d_{k,0}=\frac1{2^k}.
\ee
Using this result for $d_{k,0}$, we sum the terms of the form $(p^{2(n-k)}\ot1)(p^{2k}\ot p_\m)=p^{2n}\ot p_\m$
\be \lb{ssc}
\sum_{n=0}^\infty \sum_{k=0}^n
\b^n \frac{(-1)^{k+n}}{k!} A_{n,k} \frac1{2^k}p^{2n}\ot p_\m= e^{\frac12\ln(1+\b p^2)}\ot p_\m = \sqrt{1+\b p^2} \ot p_\m,
\ee
which agrees with the corresponding term in $\D p_\m$ \eqref{coprod}. Proceeding similarly, using the result for $c_{k-1,1}$ in \eqref{ccd},
we get \blue{$p_\m\,\frac{1+\b p^2 - \sqrt{1+\b p^2}}{p^2}p_\a\ot p^\a$}, in accordance with the corresponding term in $\D p_\m$ \eqref{coprod}.
The complete inductive proof for $\D p_\m$ \eqref{coprod} and $\cF$ \eqref{fsn} will be given elsewhere.
}

\red{As a consistency check, using the twist} \eqref{fsn} to calculate the coproduct of $M_{\mu \nu}$ we can also verify that the coproduct of the Lorentz
generators is undeformed to all orders i.e.,
\begin{equation}
\Delta M_{\mu\nu} =\Delta_0 M_{\mu \nu}.
\end{equation}

Note that \red{the} twist corresponding to the Snyder realization can be written in terms of the dilatation $D=x\bdot p$ and of $p^2$,
in a form which slightly differs from eq.\ \eqref{fsn} but is equivalent to it.

\section{Twist for the Maggiore realisation}

The same procedure can be performed for the Maggiore realisation \eqref{magg}. The coproduct, when expanded up to the third order, takes the following form
\begin{eqnarray}
\Delta p_\mu &=& p_\mu \otimes 1+1\otimes p_\mu -\frac{\b}{2} \left( p_\mu p_\alpha \otimes p^\alpha +p_\mu \otimes p^2 \right) \\
&\quad & -\frac{\b^2}{8} \left( p_\mu \otimes p^4 +p_\mu p_\alpha p^2\otimes p^\alpha \right)
-\frac{\b^3}{16} \left( p_\mu \otimes p^6 +p_\mu p_\alpha p^4 \otimes p^\alpha \right) + O(\b^4) \nonumber
\end{eqnarray}
Using the same procedure as in the previous section, we find
\begin{eqnarray}
f_1 &=& \frac{i\beta}{2} \left( p_\alpha \otimes x^\alpha p^2 +p_\alpha p_\beta \otimes x^\alpha p^\beta \right), \\
f_2 &=& \frac{i\beta^2}{8} \Bigl( p_\alpha \otimes x^\alpha p^4 +p_\alpha p^2 \otimes x^\alpha p^2 +
2p_\alpha p_\beta p_\gamma \otimes x^\alpha p^\beta p^\gamma \nonumber \\
&\quad & +2p_\alpha p_\beta \otimes x^\alpha p^\beta p^2 +2 p_\alpha p_\beta p^2 \otimes x^\alpha p^\beta \Bigr), \nonumber \\
f_3 &=& \frac{i\beta^3}{8} \Bigr( \frac{1}{2} p_\alpha \otimes x^\alpha p^6 + \frac{4}{3} p_\alpha p_\beta p^4 \otimes x^\alpha p^\beta
+\frac{3}{2} p_\alpha p_\beta \otimes x^\alpha p^\beta p^4 \nonumber \\
&\quad & +\frac{7}{12} p_\alpha p^4 \otimes x^\alpha p^2 +\frac{5}{12} p_\alpha p^2 \otimes x^\alpha p^4
+\frac{7}{3} p_\alpha p_\beta p_\gamma \otimes x^\alpha p^\beta p^\gamma p^2 \nonumber \\
&\quad & +\frac{5}{3} p_\alpha p_\beta p^2 \otimes x^\alpha p^\beta p^2 +\frac{4}{3}
p_\alpha p_\beta p_\gamma p_\delta \otimes x^\alpha p^\beta p^\gamma p^\delta
+2p_\alpha p_\beta p_\gamma p^2 \otimes x^\alpha p^\beta p^\gamma \Bigr). \nonumber
\end{eqnarray}

In this case, we were not able to obtain a closed form for the twist. However, the perturbative result, when used to calculate the coproduct of
$M_{\mu\nu}$, gives again the primitive coproduct.


\section{Conclusions}
In this paper we have investigated the most general realisations of the Snyder model compatible with undeformed Lorentz invariance,
and have calculated the twist and the R-matrix for the generic case, at leading order in the deformation parameters.
In particular, in the specific case of the Snyder realisation we were able to obtain \blue{an analytic} expression for the twist.

Our results can be rephrased using the formalism of Hopf algebroids \red{\cite{JMS, Lu, Xu, PLA2013, SIGMA2014, PLB2015, LMP2017, MS1605, JPA2017}}, that is for some
aspects more suitable for the description of the Snyder models than the usual one based on Hopf algebras, since it
deals with the full phase space. We leave however this subject to future investigations.

The results obtained in this paper may be important for the construction of a complete QFT on Snyder spaces. Some basic attempts
in this direction have been put forward in refs.\ \cite{BM2,GL,MMMS}.

\section*{Acknowledgements}
We wish to thank J. Lukierski and D. Pikuti\'c for comments and discussions.
The work of S. Meljanac has been supported by Croatian Science Foundation under the project IP-2014-09-9582,
as well as by H2020 Twinning project no 692194 "RBI-T-WINNING".
S. Mignemi wishes to thank Rudjer Bo\v skovi\'c Institute for hospitality during the preparation of this work.


\begin{thebibliography}{0}

\bibitem{Snyder} H.S. Snyder, \PR{71}, 38 (1947).
\bibitem{DFR} S. Doplicher, K. Fredenhagen and J.E. Roberts, \PL{B331}, 39 (1994).
\bibitem{Moyal} V.P. Nair and A.P. Polychronakos, \PL{B505}, 267 (2001);
 L. \red{Mezincescu}, {\tt hep-th}/0007046.
\bibitem{kappa} J. Lukierski, H. Ruegg, A. Novicki and V.N. Tolstoi, \PL{B264}, 331 (1991);
J. Lukierski, A. Novicki and H. Ruegg, \PL{B293}, 344 (1992).
\bibitem{Majid} S. Majid, {\it Foundation of quantum group theory}, Cambridge University Press 1995.
\bibitem{BM1} M.V. Battisti and S. Meljanac, \PR{D79}, 067505 (2009).
\bibitem{BM2} M.V. Battisti and S. Meljanac, \PR{D82}, 024028 (2010).
\bibitem{kappaS} S. Meljanac, D. Meljanac, A. Samsarov and M. Stoji\'{c}, \MPL{A25}, 579 (2010); \PR{D83}, 065009 (2011).
\bibitem{GL} F. Girelli and E.L. Livine, \JHEP{1103}, 132 (2011).
\bibitem{MMMS} S. Meljanac, D. Meljanac, S. Mignemi and R. \v Strajn, \red{\PL{B768}, 321-325 (2017);}
\red{S. Meljanac, S. Mignemi, J. Trampeti\'c, J. You, \arx{1703.10851}}.
\bibitem{CM} L.N. Chang, D. Mini\'c, N. Okamura and T. Takeuchi, \PR{D65}, 125027 (2002);
S. Benczik, L.N. Chang, D. Mini\'c, N. Okamura, S. Rayyan and T. Takeuchi, \PR{D66}, 026003 (2002).
\bibitem{mi} S. Mignemi, \PR{D84}, 025021 (2011); S. Mignemi and R. \v Strajn, \PR{D90}, 044019 (2014);
B. Iveti\'c, S. Mignemi and A. Samsarov, \PR{A93}, 032109 (2016).
\bibitem{LS} Lei Lu and A. Stern, \NP{B854}, 894 (2011); \NP{B860}, 186 (2012).
\bibitem{JMS} T. Juric, S. Meljanac and R. \v Strajn, \IJMP{A29}, 145022 (2014).
\blue{
\bibitem{Lu} J.-H. Lu, Internat. J. Math. {\bf7}, 47-70 (1996).
\bibitem{Xu} P. Xu, \CMP{216}, 539-581 (2001).
\bibitem{PLA2013} T. Juri\'c, S. Meljanac, R \v{S}trajn, \PL{A377}, 2472-2476 (2013).
\bibitem{SIGMA2014} T. Juri\'c, D. Kova\v{c}evi\'c, S. Meljanac, SIGMA {\bf10}, 106 (2014).
\bibitem{PLB2015} J. Lukierski, Z. \v{S}koda, M. Woronowicz, \PL{B750}, 401-406 (2015).
\bibitem{LMP2017} S. Meljanac, Z. \v{S}koda, M. Stoji\'c, Lett. Math. Phys. {\bf107}, no.3, 475-503 (2017).
\bibitem{MS1605} S. Meljanac, Z. \v{S}koda, \arx{1605.01376}.
\bibitem{JPA2017} S. Meljanac, D. Meljanac, A. Pachol, D. Pikuti\'c, \JoP{A50}, no.26, 265201 (2017).
}
\bibitem{Mag} M. Maggiore, \PL{B319}, 83 (1993).
\bibitem{Svrtan} S. Meljanac, Z. \v{S}koda and D. Svrtan, SIGMA {\bf8}, 013 (2012).
\bibitem{KM} D. Kovacevic and S. Meljanac, \JoP{A45}, 135208 (2012).
\bibitem{KMSS} D. Kovacevic and S. Meljanac, A. Samsarov and Z. \v Skoda, \IJMP{A30}, 1550019 (2015).
\bibitem{JMP} T. Juric, S. Meljanac and D. Pikutic, \EPJ{C75}, 528 (2015).
\bibitem{MM} S. Meljanac, D. Meljanac, F. Mercati and D. Pikutic, \PL{B766}, 181 (2017).
\bibitem{GGH} T. R. Govindarajan, K.S. Gupta, E. Harikumar, S. Meljanac and D. Meljanac, \PR{D77}, 105010 (2008).
\bibitem{MSS} S. Meljanac, A. Samsarov and R. \v Strajn, \JHEP{1208}, 127 (2012).
\bibitem{GM}  G. Gubitosi and F. Mercati, \CQG{20}, 145002 (2013).
\bibitem {maj} S. Majid, {\it Algebraic approach to quantum gravity II: noncommutative spacetime},
in D. Oriti, {\it Approaches to quantum gravity}, Cambridge University Press 2009.

\end{thebibliography}
\end{document}